\documentclass[11pt]{article}
\usepackage{jheppub}
\pdfoutput=1
\usepackage{simeontex-csm}

\usepackage{tikz}
\usepackage{trsym}
\usetikzlibrary{arrows,decorations.markings}
\usetikzlibrary{calc}

\usepackage{braket}

\usepackage{hyperref}
\hypersetup{colorlinks=true,
linkcolor=magenta,          % color of internal links (change box color with linkbordercolor)
citecolor=magenta,        % color of links to bibliography
filecolor=red,}
\definecolor{refkey}{rgb}{0.85, 0.0, 0.3}
\definecolor{labelkey}{rgb}{0.2,0.2,0.6}

\newcommand{\sugi}[1]{\textcolor{blue}{(SS: #1)}} %% to add a comment
\allowdisplaybreaks[1]

\def\pitem#1{\purple{\item{{#1}}}}

\makeatletter
  \def\Tr{\mathop{\operator@font tr}\nolimits}
  \makeatother
\def\trB{\mathop{\Tr_B}}
\def\trA{\mathop{\Tr_A}}
\newcommand{\OfficialTitle}{Page Curves for General Interacting Systems}
\hypersetup{pdfauthor={Hiroyuki Fujita and Yuya Nakagawa and
     Sho Sugiura and Masataka Watanabe},pdftitle={\OfficialTitle}}
\title{\OfficialTitle}

\author[1,2]{Hiroyuki Fujita}
\author[1,2]{Yuya O. Nakagawa}
\author[3]{\par Sho Sugiura}
\author[2,4]{Masataka Watanabe}
\affiliation[1]{\small Institute for Solid State Physics, The University of Tokyo. Kashiwa, Chiba 277-8581, Japan}
\affiliation[2]{\small Department of Physics, Faculty of Science, The University of Tokyo, Bunkyo-ku, Tokyo 133-0022, Japan}
\affiliation[3]{\small Department of Physics, Harvard University}
\affiliation[4]{\small Kavli Institute for the Physics and Mathematics of the Universe (WPI),The University of Tokyo Institutes for Advanced Study, The University of Tokyo, Kashiwa, Chiba 277-8583, Japan}

\emailAdd{masataka.watanabe@ipmu.jp}

\date{}
\normalsize

\abstract{
\normalfont \noindent

We calculate in detail the Renyi entanglement entropies of cTPQ states as a function of subsystem volume, filling the details of our prior work \hyperlink{name}{[Nature Communications 9, 1635 (2018)]}, where the formulas were first presented. Working in a limit of large total volume, we find universal formulas for the Renyi entanglement entropies in a region where the subsystem volume is comparable to that of the total system. The formulas are applicable to the infinite temperature limit as well as general interacting systems. For example we find that the second Renyi entropy of cTPQ states in terms of subsystem volume is written universally up to two constants, $S_2(\ell)=-\ln K(\beta)+\ell\ln a(\beta)-\ln\left(1+a(\beta)^{-L+2\ell}\right)$, where $L$ is the total volume of the system and $a$ and $K$ are two undetermined constants. The uses of the formulas were already presented in our prior work and we mostly concentrate on the theoretical aspect of the formulas themselves. Aside from deriving the formulas for the Renyi Page curves, the expression for the von Neumann Page curve is also derived, which was not presented in our previous work.

      }

\begin{document}
\maketitle 
      \newpage
%\restoregeometry
\setcounter{tocdepth}{2}
\setcounter{secnumdepth}{4}
%\tableofcontents
%\newpage

\section{Introduction}
\label{sec:intro}

\if0
\todo[inline]{Add something nice that attracts people.
Add some killer words. Ask Sho about the motivation of the work.}
Entanglement has become super-popular recently both in high-energy physics and
condensed matter physics.
Compared with the ground states, the structure of entanglement in finite temperature is not
so investigated.
It is important to understand the finite-temperature states because it gives the information on
the thermal entropy, the central charge(?), etc.

The notion of entanglement has become quite popular these days
as a common language over physicists in the fields of high-energy, condensed matter,
and quantum information. 
One useful measure of entanglement is the
entanglement entropy,
which scales the quantum correlation of one subsystem with the other.
The entanglement entropy of the ground state of the system
is known to obey the area-law
\fi

The notion of entanglement has become popular these days
as a common language over physicists in the fields of high-energy, condensed matter,
and quantum information \cite{Ryu:2006bv,KitaevPreskill2006, LevinWen2006,Hawking_paradox}. 
One useful measure of entanglement is the
entanglement entropy,
which quantifies the quantum correlation of one subsystem with its compliment.
The entanglement entropy of the ground state of locally interacting systems
is known to obey an area-law. 
%,
%where the quantity grows as
%$O(C^{d-1})+\text{$\log$-corrections}$, where $C$ is a typical length scale of the subsystem and $d$ is the spatial dimension of the system (for a review see \cite{Eisert2010} and references therein). 
%Incidentally, this has an explanation in terms of the gravity dual of the theory, called Ryu-Takayanagi formula \cite{Ryu:2006bv, Hubeny:2007xt, Lewkowycz:2013nqa}.
%%$O(C^{d-1})+\text{$\log$-corrections}$ where $C$ is the length of the boundary of the subsystem and $d$ is the spatial dimension of the system. 
The entanglement entropy
of the pure quantum state which have large amplitudes of excitations, however, behaves differently;  
when the subsystem volume is small compared with the total volume, it follows a volume-law,
meaning the entropy grows in proportion to the subsystem volume, i.e., grows as
$O(C^d)$ \cite{Takayanagi2010}.
%This is roughly because one needs to specify the
%field configuration of the whole space, as compared with
%the entangling surface only for ground states, to determine the
%entanglement degrees of freedom.
%In other words, 
This is roughly because one has to take the thermal entropy of the subsystem itself
into account with respect to the excited states.
Therefore, at small subsystem sizes, the thermal effect evades the quantum effect.

How does the entanglement entropy of excited states behave when
the subsystem volume is not necessarily small --
Can one recover the information about quantum effects in that way?
In particular, what will be the deviation from %/the correction to 
the volume law
when the subsystem volume is almost half the total volume of the system? 
These are the questions to be answered in this paper. 
These questions are very much worth asking 
as their answer should fully characterise the entanglement
entropy for any subsystem sizes, in comparison to the ``volume-law'',
which is a statement about the entanglement entropy for small subsystem sizes and only teaches us the thermal information about the system. 
To answer these questions, we have to
calculate the entanglement entropy against subsystem volume
as a functional form -- this graph is called the 
von Neumann/$n$-th Renyi
Page curve for von Neumann/$n$-th Renyi entanglement entropy. 
The Page curve is calculated both in the context of Black Hole formation/evaporation \cite{Takayanagi:2010wp} and in the context of the foundation of quantum statistical mechanics \cite{Calabrese2004,Calabrese_quench,PhysRevX.8.021026}.
The 2nd Renyi Page curve is even observed in experiments using ultra-cold atoms \cite{Science2016}. 

Although these observations are limited to specific models, generically Page curves share several common features. 
%A similar structure for states after a quench was observed in the context of Black Hole formation/evaporation in \cite{Takayanagi:2010wp}.
The entanglement entropy scales linearly in proportional the volume of the subsystem as far as the subsystem is sufficiently smaller than the entire system. 
%Note that from here one can recover the volume-law for the random spin system, whose coefficient is given by $\ln 2$.
%This is 
%just the thermal entropy of the system at infinite temperature.
%Also, 
%You can see that 
When the volume of the subsystem is comparable to that of the entire system, the entanglement entropy deviates from the above volume law. 
It starts decreasing when the subsystem volume is larger than half the total volume, and eventually vanishes when the subsystem is as large as the entire system (See Fig. \ref{a} for a similar plot for the second Renyi entropy). 
In this paper, we present the results which reproduce these features without restricting a Hamiltonian to a specific model. 
This result is important as it could qualitatively explain the
Black Hole information paradox, considering the subsystem as 
Black Hole radiation and the compliment as the remaining Black Hole \cite{Almheiri:2012rt,Harlow:2014yka,HottaSugita}.
Therefore, it will be of great interest to extract universal information about Page curves, irrespective of the choice of a particular model. 

At infinite temperature $\beta=0$, the equilibrium state $e^{-\beta H}/ \tr[e^{-\beta H}]$ becomes independent of the Hamiltonian. 
%Since the equilibrium state is independent of the Hamiltonian at $\beta=0$, the Page curve at infinite temperature is trivially universal. 
Similarly, the Page curve at infinite temperature is trivially universal at $\beta=0$. 
In the monumental work \cite{Page:1993df} published in 1993, Don N. Page derived the von Neumann Page curve of random spin-$1/2$ systems:
\begin{equation}
S(\ell)=\ell\ln 2-\frac{1}{2}\times \frac{2^\ell}{2^{L-\ell}},
\label{page}
\end{equation}
where $L$ and $\ell$ is the number of total spins and 
the number of spins the subsystem contains, respectively. 
As the Hamiltonian of the random spin system is given by $H=0$,
this gives the form of the Page curve for any systems at
infinite temperature ($\beta=0$). 
\eqref{page} indeed reproduces the above-mentioned features of the Page curve. 
However, all the model dependence is smeared out at infinite temperature. 

In order to study the Page curve at finite temperatures, we need corresponding pure quantum states. 
\if0
Still, one has to consider finite-temperature systems to sharpen the understanding towards Black Hole information paradox.
More importantly, but related, the entanglement entropy of excited states also appears in the context of thermalization \cite{ETHDeutch1991, ETHSrednicki1994, Thermalization2008,PolkovnikovReview2011,Biroli2010,Iyoda2017}. 
A quantum pure state in a scrambled system thermalises using its own subsystem as a thermal bath, and
the quantum entanglement substitutes the role of the thermodynamic entropy \cite{Popescu2006,Goldstein2006,Ruihua2016}
as it reaches the equilibrium.
Such an effect and the system's entanglement/Renyi entropy is already observed in experiments using ultra-cold atoms \cite{Science2016}. 
This is the reason why the theory is needed which can generally characterize the subsystem volume-dependence of the Page curve in general interacting systems which are fast-scrambled. 
In general interacting systems, however, it is usually difficult to perform a generic computation applicable to a large class of theories.
Even in this case where we restrict attention to fast-scrambled systems, they usually lack simple universal characterizations.
\fi
%The characterization is believed to be made by using out-of-time-order-correlators (OTOCs) \cite{Maldacena:2015waa}, but it still is difficult enough to apply such diagnosis to the problem in question. 
One candidate is the thermal pure quantum (TPQ) state \cite{Sugiura2012, Sugiura2013}. 
In the context of the foundation of quantum statistical mechanics, a quantum pure state in a scrambled system is believed to thermalise using its own subsystem as a thermal bath \cite{ETHDeutch1991, ETHSrednicki1994, Thermalization2008,PolkovnikovReview2011,Biroli2010,Iyoda2017}, and 
the quantum entanglement takes over the role of the thermodynamic entropy \cite{PhysRevX.8.021026,Anatoly2016,HottaSugita}. 
The TPQ states, which are a set of typical random pure states, are the state which mimics a pure state after the thermalisation. 
Specifically, we can prove that the expectation values of any local operators 
distribute around thermally averaged values of those operators,
with their variances exponentially small as the total volume of the system grows \cite{Sugiura2012, Sugiura2013}. 
One advantage of this method is that it is computationally easy to extract information about physical observables.
The expectation value can just be extracted by averaging over random variables, or further, if you pick one random state in a collection of cTPQ states, the value of an observable you get is exponentially close to the one you might have got for the thermal expectation value of the observable.
The TPQ states serve as a tool in analyzing a system after relaxation to the thermal equilibrium. 

\if0
TPQ states are a collection of pure states weighed by random variables, and one takes average over them when computing physical observables.
These states can in principle retain the pureness of the quantum state as well as correctly reproduce the local physics, as they are 
known to correctly reproduce the thermal expectation values of local observables.
For these reason, the generalisation of \cite{Page:1993df} to finite temperature should be made using TPQ states.
Note that there are two classes of TPQ states, canonical and micro-canonical type, and
our previous paper \cite{Fujita:2017pju} used the former, while \cite{TChengTarun} (appeared on the same day as \cite{Fujita:2017pju}) used the micro-canonical type.\footnote{Incidentally in our previous paper the above motivation was partly satisfied -- there a simple diagnosis for fast-scrambled systems were proposed, such that
the entanglement entropy of those systems are well-fitted by our formula derived using TPQ states.}
\fi

Considering the above situation,
in this paper, we set out to compute the calculation of the entanglement entropy using canonical thermal pure quantum (cTPQ) states.\footnote{There are two classes of TPQ states, canonical and micro-canonical type, and
our previous paper \cite{Fujita:2017pju} and this work uses the former, while \cite{TChengTarun} (appeared on the same day as \cite{Fujita:2017pju}) used the micro-canonical type. The difference between them is the existence of the energy variance; the energy variance of the former is $O(\sqrt{L}$) while that of the latter is $O(1)$. Like the ensemble of the statistical mechanics, one should choose appropriate TPQ state depending on a situation.} 
%tackle the generalisation of \cite{Page:1993df}
%to finite temperature using thermal pure quantum (TPQ) states.
%These states, first introduced in \cite{Sugiura2012}, are
%known to be a set of typical random states with same macroscopic variables. 
%The expectation values of any local operators 
%distribute around thermally averaged values of those operators,
%with their variances exponentially small as the total volume of the system grows. 
We will first try to expand \cite{Page:1993df} and calculate the
$n$-th Renyi Page curve of the random spin system, and then
compute it for general interacting systems at finite temperature.
Especially the second Renyi Page curve for general interacting systems
and prove that it behaves universally up to two constants (one for the offset
of the entropy, and the other for the slope of the volume-law).
We also compute the von Neumann Page curve by taking a limit of the Renyi index $n\to 1$.\footnote{There also appeared a paper \cite{HUANG2018} which derives the von Neumann entanglement entropy of chaotic systems analytically. The result there is also conjectured to be universal, and clearly is complimentary to our result about Renyi.}
The readers are refereed to our previous work \cite{Fujita:2017pju} for uses and numerical evidences that back up this result -- we conjectured that the preciseness of the fit of our formula to the actual Page curve constitutes
the diagnosis for fast-scrambled systems.

The plan of the paper is as follows. In Section \ref{notation}, we fix the notation
and briefly review some of the properties of Page curves.
In Section \ref{sec:nthrenyi} we calculate
the $n$-th Renyi Page curves of the random spin system for any $n$ and
prove that the von Neumann Page curve, obtained by taking $n\to 1$ matches with the previous result by Page, (\ref{page}) \cite{Page:1993df}.
In Section \ref{fintemp},
we expand the previous section's result to general interacting systems 
at finite temperature using TPQ states.
We especially focus on the second Renyi Page curve and stress
that its form is determined by two constants which can be fitted with numerical data.
We also compute the von Neumann Page curve by taking a limit of $n\to 1$.
In Section \ref{Example}, we present an example to back up our formula.

\section{Notations and properties of Page curves}
\label{notation}

\subsection{Definitions and notations}

Let us consider a general lattice system $\Sigma$ with $L$ spins. 
%We only consider $s=1/2$ spins in this paper for simplicity,
%so that the dimension of the Hilbert space $\mathcal{H}_\Sigma$ of $\Sigma$
%is $d\equiv 2^L$ -- extensions of our result to any other  spin degrees of freedom is just arithmetic
%and we do not care to mention.
Let us now divide $\Sigma$ into two parts, $A$ and $B$, to
evaluate the entanglement of the system.
We set the number of spins in $A$ and $B$ to be $\ell$ and $m$, respectively,
and denote the dimension of each Hilbert space associated with
$A$ and $B$ as $d_A\equiv s^\ell$ and $d_B\equiv s^m$ where $s$ is the degree of freedom of each spin. 
%$A$ and $B$ as $d_A\equiv 2^\ell$ and $d_B\equiv 2^m$.
Note that $L=\ell+m$, so that the dimension of the Hilbert space $d$ is $d=d_Ad_B$.

By using the notations above, 
%the density matrix of the system
%associated with a wave function $\Ket{\psi}$ 
%is defined as $\rho\equiv\Ket{\psi}
%\Bra{\psi}$, and 
the reduced density matrix on subsystem $A$ constructed
from the density matrix $\rho$ on $\Sigma$ is defined as
\begin{equation}
\rho_A\equiv\Tr_B \rho,
\end{equation}
and by using this, the $n$-th Renyi entanglement entropy is defined via
\begin{equation}
S_n^A\equiv\frac{1}{1-n}\ln \lrd \trA \rho_A^n \rrd.
\label{nthrenyi}
\end{equation}
The von Neumann entanglement entropy is defined as
\begin{equation}
S^A\equiv-\trA(\rho_A\ln\rho_A),
\end{equation}
and can be calculated by performing an analytic continuation of $S_n$
and by taking $n\to 1$:
\begin{equation}
S^A\equiv S_1^A\equiv \lim_{n\to 1}S_n^A
\end{equation}

\subsection{(Generalised) Page curves}

A Page curve, originally introduced in \cite{Page:1993df} is a function of entanglement entropy for the random spin system in terms of subsystem volume.
Here, we generalise the concept of it to general interacting systems:
A Page curve is a graph
of entanglement entropy plotted against the subsystem volume, $\ell$.
The entanglement entropy is indeed shape-dependent in general \cite{Faulkner:2015csl},
but understand this statement as we have agreed upon one way
of choosing the subregion shape.

We hereafter call $S_n^A(\ell)$ as the $n$-th Renyi Page curve
and $S^A(\ell)=S^A_1(\ell)$ as the von Neumann Page curve.
Additionally, note that $S^B(\ell)$ denotes the entanglement entropy
traced over $A$ (i.e., $S^B=-\trB(\rho_B\ln\rho_B)$) when $B$ contains $\ell$ spins.
Here after we will omit the superscript $A$ when there are no confusions.

One of the important properties of the original Page curve is the symmetry under subregion-subregion interchange, i.e.,
$S_n^A(\ell)=S_n^A(L-\ell)$.
This is directly inherited to generalised Page curves, if the density matrix in question is pure.
This will be important in the following sections as we study them in more details.

\section{Calculation of the entanglement entropy for the random spin system}
\label{sec:nthrenyi}
Calculation of the von Neumann (entanglement) entropy 
and the second Renyi (entanglement) entropy of the random spin system
is
already done in \cite{Page:1993df} and \cite{doi:10.1143/JPSJ.74.1883}.
We mainly follow the latter work to expand this calculation
to $n$-th Renyi entropy. We will also check if this result is consistent with
the von Neumann entropy given in the former.

\subsection{Calculation of the $n$-th Renyi entropy}
\label{ssec:nthrenyi}
\if0
\textit{Although the authors derived them independently, the results of Sec. \ref{ssec:nthrenyi} was already derived in the context of Random Matrix Theory. The readers are refereed to \cite{Dumitriu03eigenvaluestatistics} and references therein for more information.
Sec. \ref{new} and onwards the content is completely new.}
\fi
\subsubsection{Random pure state}
Let us consider the spin system $\Sigma$ with $L$ random spins.
We divide the system up into two pieces as in Section \ref{notation}.
\if0
The dimension of the Hilbert space $\mathcal{H}_\Sigma$ of $\Sigma$
is then $d\equiv 2^L$.
Let us now divide $\Sigma$ into two parts, $A$ and $B$, to
evaluate the entanglement of the system.
We set the number of spins in $A$ and $B$ to be $\ell$ and $m$, respectively,
and denote the dimension of each Hilbert space associated with
$A$ and $B$ as $d_A\equiv 2^\ell$ and $d_B\equiv 2^m$.
Note that $L=\ell+m$, so that $d=d_Ad_B$.
\fi
Following the notations there, general wavefunctions of the system
can now be written as
\begin{equation}
\Ket{\psi}=\sum_{a,b}c_{a,b}\Ket{a}\otimes\Ket{b},
\label{wf}
\end{equation}
%\todo{Not sure if this notation is clear enough.}
%I want people to be sure that $a$ and $n$ run over different subspaces.
We call this a random pure state, where we take $c_{*,*}$ to be uniformly distributed on a unit sphere in $\mathbb{C}^d$.

\subsubsection{Calculation of $\overline{\trA \rho_A^n}$}
\label{calcrand}
%\subsubsection{Diagrammatic representation of $\overline{\trA \rho_A^n}$}

%\paragraph{Diagrammatic representation}
By straightforward calculation, we obtain
\begin{equation}
\trA\rho_A^n=\sum_{a_*,b_*}c^*_{a_1b_1}c_{a_2b_1}
c^*_{a_2b_2}c_{a_3b_2}\cdots c^*_{a_nb_n}c_{a_1b_n},
\end{equation}
whose cyclicity of the index we represent by the diagram below:
\begin{equation}
\includegraphics[width=0.3\columnwidth]{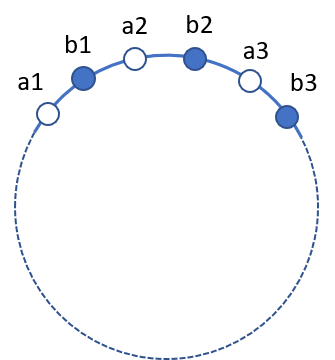}
\label{npolygon}
\end{equation}

\begin{figure}[htbp]
        \begin{center}
           \includegraphics[width=0.7\columnwidth]{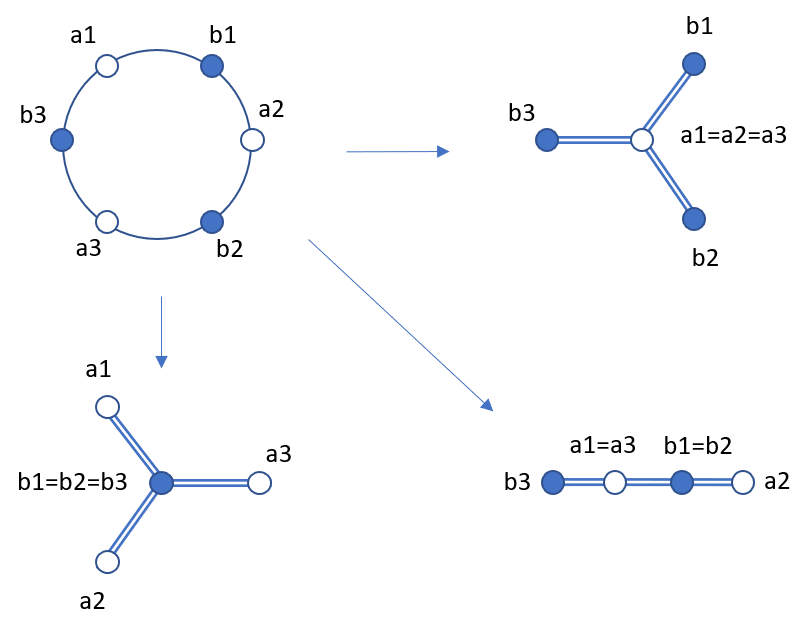}
    \caption{Fig. \ref{2}: All the $n=3$ graphs at leading order in large-$d$.}
    \label{2}
        \end{center}
        %\caption{Fig. \ref{2}: Ways to contract $n=3$ graph.}
    \end{figure}

We now try to compute $\overline{\trA \rho_A^n}$.\footnote{The result seems to have been known already in a completely different context of Random Matrix Theory \cite{Dumitriu03eigenvaluestatistics}, but for the sake of the discussion in the next section, let us reproduce the result in a different way, using diagrammatic approach.}
Note that $\ln \overline{\trA \rho_A^n}$ is the same as computing $\overline{\ln \trA \rho_A^n}$ at leading order in large-$d$.
The complete proof of this fact as well as intuitive explanation is given in Sec. \ref{referee}.
Because of the results shown in Appendix. \ref{ssec:rmt},
non-vanishing contributions after averaging
are represented by diagrams made by joining $\TransformHoriz $ together in
(\ref{npolygon}), meaning two pairs of indices, $(a,b)$'s,
are the same.  
We show an example of this contraction for $n=3$ in Fig. \ref{2}
\if0
We analyse the result above for two cases -- (a) for 
$\ell=m$ and (b) for $\ell<<m$

\paragraph{(a) $\ell=m$}
Let us analyse the result when $\ell$ is equal to $m$, i.e., when
$A$ is half the total space.
In the region where $\ell$ and $m$ are much greater than 1,
at leading order in $d$-scaling only relevant contractions
of the graph are such that
we contract every link just once and that
there is no loop in the resulting diagram.
\fi

%\paragraph{Narayana numbers}
In a region where $2^\ell$ and $2^m$ are much greater than 1 (note that $2^\ell/2^m$ could be of order 1),
at leading order in $d$-scaling only relevant contractions
of the graph are such that
we contract every link just once and that
there is no loop in the resulting diagram.
The contribution from one resulting diagram will 
be equal to
$d_A^{n_A}d_B^{n_B}\times \overline{|c|^2\cdots|c|^2}$ when the
resulting number of white and blue dots, respectively, is
$n_A$ and $n_B$, where $n_A+n_B=n+1$.
We hereafter call those diagrams as diagrams of the order $n_A$.

\begin{center}
    \begin{figure}[h]
        \begin{center}
            {%\small
            \resizebox{0.7\linewidth}{!}{\input{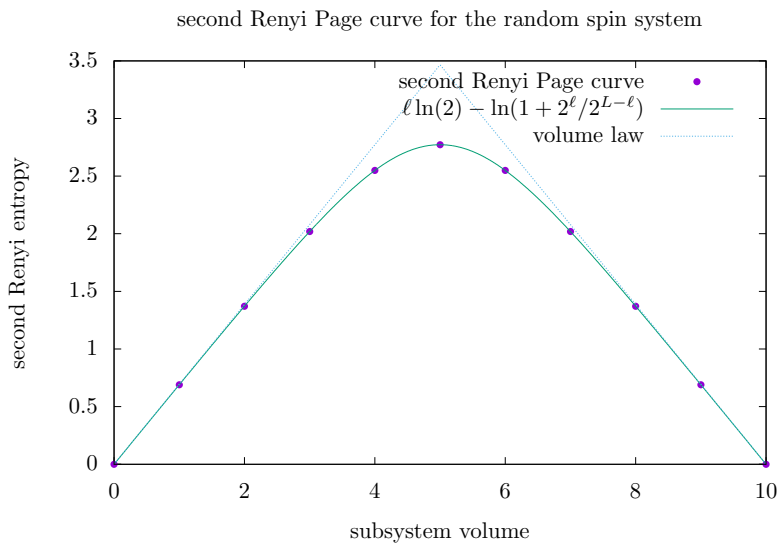}}
            }
            \caption{Fig. \ref{a}: The second Renyi Page curve for the random spin system.
            The curve is convex and symmetric at the centre.
            }
            \label{a}
        \end{center}
    \end{figure}
\end{center}

Now, what is the number of diagrams of the order $n_A$ for 
general $n$ and $n_A$?
This number is the same number as you might have got if you counted the
number of non-crossing partitions of $\{1, 2, \dots, n\}$ of the rank 
$n_A$, meaning you divide them up into non-crossing $n_A$ pieces. 
This number is already known as Narayana number \cite{2012arXiv1206.0803B}, denoted and defined by $N(n,n_A)\equiv\dfrac{1}{n}\dbinom {n}{n_A} \dbinom {n}{n_A-1}$.
We get, by using this notation, the following;
\begin{equation}
\overline{\trA \rho_A^n}
=\sum_{\text{All}}(\text{diagrams})
=d_A^{1-n}\times\sum_{k=1}^{n} N(n,k)\left(\frac{d_A}{d_B}\right)^{k-1},
\label{trace}
\end{equation}
and the $n$-th Renyi entropy of the random spin system becomes
\begin{align}
S_n&=\ln{d_A}-\frac{1}{n-1}\ln\left[{\sum_{k=1}^{n} N(n,k)\left(\frac{d_A}{d_B}\right)^{k-1}}\right]\\
&=\ell\ln{2}-\frac{1}{n-1}\ln\left[{\sum_{k=1}^{n} N(n,k)\left(\frac{d_A}{d_B}\right)^{k-1}}\right].
\label{vollaw}
\end{align}
This means that the $n$-th Renyi entropy of the random spin system approximately
follows a volume law ($\ell\ln 2$) when the subsystem $A$ is small, which
is then rounded off by the second term as $A$ gets bigger.
Especially when $A$ makes up half the volume of the total system, i.e., 
when $d_A=d_B$, we have 
\begin{align}
S_n(\ell=L/2)&=
\frac{L}{2}\ln 2-\frac{1}{n-1}\ln\left[{\sum_{k=1}^{n} N(n,k)}\right]\\
&=\frac{L}{2}\ln 2-\frac{1}{n-1}\ln\left[C(n)\right],
\label{nthrenyimaximal}
\end{align}
where $C(n)$ is the Catalan number \cite{2012arXiv1206.0803B}, defined by $C(n)=\dfrac{1}{n+1}
\dbinom {2n}{n}$.
We show a graph of the second Renyi Page curve in Fig. \ref{a}.
%\todo[inline]{Graphs!}

\subsection{Sanity check: Analytic continuation to $n=1$}
\label{new}
After getting the results for Renyi entropies for general integer $n$,
everyone should be naturally tempted to look into von Neumann entropy by
performing an analytic continuation to $n=1$.
We are going to first see the maximal value of the
von Neumann entropy for simplicity, and then determine the
whole functional form of the entanglement entropy to see
if it really matches the result given in \cite{Page:1993df}.
\subsubsection{Entanglement entropy at its maximal value}
Entanglement entropy, von Neumann or Renyi,
takes its maximal value when subsystem $A$ makes up half the volume
of the total system. Looking at (\ref{nthrenyimaximal})
and performing an analytic continuation, we get
the maximal value of the von Neumann entropy achieved at 
$\ell=L/2$:
\begin{equation}
S(\ell=L/2)=\frac{L}{2}\ln 2-\lim_{n\to 1}\frac{\ln{C(n)}}{n-1}=\frac{L}{2}\ln 2-\frac{1}{2}
\end{equation}
We can also see with ease that
\begin{equation}
S(\ell=0)=0
\end{equation}
These perfectly matches the prediction made in \cite{Page:1993df}.

\subsubsection{Analytic continuation of the whole function}

Analytically continuing the Page curve to the von Neumann Page curve is difficult, but can be done using the knowledge of special functions.
The actual computation is given in the Appendix, and we only present the result here;
\begin{equation}
S=\ell\ln 2-\frac{1}{2}\frac{d_A}{d_B},
\label{pages}
\end{equation}
This reproduces the result given by Page \cite{Page:1993df} in 1993
modulo terms that vanish at large-$d_A$ and $d_B$.
The reason for the difference of order $1/d_A$ or $1/d_B$ is explained in the next subsection.

\subsection{Aside: region where the subsystem Hilbert space dimension is small}
\label{aside}
%We would now elaborate a bit about the applicability condition of
%the above discussions.
In a region where the subsystem Hilbert space dimension is small, or
specifically, where $d=2^L\muchgreaterthan 1$ but $d_A=2^\ell=O(1)$, 
we will have to add corrections to the result above.
Since we only have $d_B$-scaling instead of $d$-scaling in that region,
we have to take into account terms with the
same number of $d_B$ but with lesser number of $d_A$. In other words
we are forced to add diagrams contracted twice or more to the above result.
The largest contributions of those, large-$d_B$-wise, are made
by contracting two white dots in graphs of order two. 
They scale as $O(d_A^{1-n}/d_B)$ in
${\overline{\trA \rho_A^n}}$,
%given in (\ref{trace}), 
and hence the $n$-th Renyi entropy will be modified like
\begin{equation}
S_n(\ell)=
%S_{n}^{0}(\ell)\times\ln\left(1+O(1/d_B)\right)=
S_{n}^{0}(\ell)+O\left(1/d_B\right),
\end{equation}
where $S_n^0$ is the right hand side of (\ref{vollaw}),
\begin{equation}
S_n^0
=\ell\ln{2}-\frac{1}{n-1}\ln\left[{\sum_{k=1}^{n} N(n,k)\left(\frac{d_A}{d_B}\right)^{k-1}}\right].
\end{equation}
The correction of order $O(1/d_B)$, therefore, is
present for the von Neumann, as well as Renyi, entropy --
this explains the $1/d_B$ discrepancy of (\ref{pages})
from the result given in \cite{Page:1993df}.

\section{Extension to finite temperature -- TPQ state
}
\label{fintemp}

\subsection{Set up and main results}
\subsubsection{TPQ state -introduction}
Now we are going to consider a general shift-invariant, interacting system with
Hamiltonian $H$ at inverse temperature $\beta$.
We are going to prepare a set of states, called thermal pure quantum (TPQ) states
\cite{Sugiura2012, Sugiura2013}, containing random variables as in (\ref{wf}),
and calculate various quantities by taking an average.
We particularly consider a cTPQ state, which is a TPQ state which corresponds to the canonical ensemble. The cTPQ state is defined in terms of Hamiltonian of the system as
\if0
\begin{equation}
\Ket{\psi}=\frac{1}{\sqrt{\mathcal{Z}}}
\sum_{a,b}c_{a,b}e^{-\beta H/2}\Ket{a,b},
\end{equation}
where 
\begin{equation}
\mathcal{Z}\equiv \sum \sum_{a,b,a',b'} \Bra{a,b} e^{-\beta H}\Ket{a,b}	
\end{equation}
\fi
\begin{equation}
\Ket{\psi}=\frac{1}{\sqrt{\Tr {\left(e^{-\beta H}\right)}}}
\sum_{a,b}c_{a,b}e^{-\beta H/2}\Ket{a,b}
\end{equation}
Note that these wave functions are \textit{not} normalised \textit{per se} --
they rather normalise to unity after being averaged over random variables, $c_{a,b}$.
The above two possible normalisations only make a subleading
difference in any of the arguments below in terms of large-$d$ scaling,
and hence for the sake of convenience we adopt the latter convention.

The most significant property of the TPQ state is that the TPQ state is a pure quantum state yet looks thermal; 
An expectation value of this state is almost equal to the corresponding ensemble average. For any few-body observable $A$, the following relation holds 
\begin{align}
	{\rm Prob}\left(\left|
	\bra{\psi} A \ket{\psi} - \Tr \left(A {e^{-\beta H}\over \Tr \left(e^{-\beta H} \right)}\right) \right| 
	\geq
	\epsilon
	\right) 
	={ e^{-O(-L)}\over \epsilon^2},
	\label{cTPQ A}
\end{align}
where ${\rm Prob}$ is a probability which are averaged over a set of random variables $c_{a,b}$. 
\eqref{cTPQ A} means that the cTPQ state is almost identical to the Gibbs state as far as we observe few-body observables. 
The concept of the pure quantum states which represent thermal equilibrium arose in the context of black hole physics and in the studies of the foundation of statistical physics independently. 
The TPQ state is a specific example of such states. 

One conceptual explanation of the cTPQ state is that it is a typical example of pure quantum states after a quantum quench and a subsequent relaxation to equilibrium. 
Suppose that we have an eigenstate $|\psi \rangle$ of a Hamiltonian $H_0$ and change the Hamiltonian to $H_1$. 
Then $|\psi \rangle$ is written in terms of the eigenstates of $H_1$.
\begin{align}
	\ket{\psi}=\sum_n {a_n} \ket{n},
\end{align}
where $\ket{n}$ is an eigenstete of $H_1$, $H_1\ket{n}=E_n\ket{n}$. After the time evolution, each eigenstate acquires a different phase.
\begin{align}
	\ket{\psi}=\sum_n {a_n} e^{-{i\over \hbar} E_n t}\ket{n}.
\end{align}
When the change of the Hamiltonian is large enough and macroscopic, the quantum state has the energy variance which is determined by thermodynamics. Namely, the distribution of the amplitudes $|a_n|^2$ should be similar to the canonical distribution. 
When the time $t$ is sufficiently large, we can approximate these phases random (for more detailed conditions, see, e.g., \cite{PReimanTimeEv2008}). 
We thus approximate that the phases is random and the amplitude $|a_n|^2$ distributes around $e^{-\beta E_n}$. 
Physically speaking, the cTPQ state mimics the energy distribution of the quantum quench and the phases of the subsequent relaxation process. 
Of course, however, since the realizations of perfect random variables $\{c_{a,b}\}$ are difficult, we should keep in mind that the cTPQ state is not valid at the microscopic level (e.g., each amplitude and phase) but valid when we look at statistical-mechanical quantities. 

Since the cTPQ state is a good example of pure states which are in equilibrium, the natural question is how much entanglement entropy this state has. 
The entanglement entropy of such pure states which look thermal is gathering attention recently, because the entanglement entropy substitutes the thermodynamic entropy in such states. 
However, quantitative calculations of the entanglement entropy of such states is limited to some specific Hamiltonians which are integrable. 
In this section, we thus calculate the entanglement entropy of the cTPQ state. 
Since we do not restrict ourselves to any specific Hamiltonian, the applicability of our result is broad; 
We numerically verified that the entanglement entropy of the cTPQ state indeed describes a generic behavior among the entanglement entropy of such states in equilibrium \cite{Fujita:2017pju}. 

We also have the TPQ state which corresponds to other ensembles, the microcanonical ensemble and the grandcanonical one. Important difference between the microcanonical TPQ state and the cTPQ state is the presence of the energy variance. The microcanonical TPQ state does not have the energy variance. It results in a different behavior of the size dependence of the entanglement entropy \cite{PhysRevX.8.021026}. 
This is explained as follows. 
When we look at a vanishingly small part of the system, the difference among the ensembles does not appear, because the rest of the system behaves as a heat bath. This is so-called the equivalence of the ensembles. 
When we look at not-vanishingly-small part of the system, however, the rest of the system cannot completely behaves as a heat bath. 
The entanglement entropy which we are interested in is in this regime. 
We should choose an appropriate TPQ state and then obtain a correct answer. 
%In analyzing black hole physics, the cTPQ is the relevant one because...

\subsubsection{Main result and its implication} \label{Main Result}
We briefly summarize our main results here. Their derivations are shown in the following sections. 
First, we can explicitly calculate the $n$-th Renyi entanglement entropy of the cTPQ state in terms of its subsystem volume. It is written as 
\begin{align}
S_2&=-\ln K(\beta)+\ell\ln a(\beta)-\ln\left(1+a(\beta)^{-L+2\ell}\right)\label{2ndRenyiMain}\\
S_3&=-\ln K_2^{\prime}(\beta)+{1\over 2}\ell\ln b(\beta)-{1\over 2}\ln\left(1+K_1'(\beta){b(\beta)^\ell \over a(\beta)^{L}}+b^{\prime}(\beta)^{-L+2\ell}\right).
\label{3rdRenyiMain}
\end{align}
%by using several unknown $O(1)$ coefficients. 
Here $K$ and $K'$ are $O(1)$ constants which depends on $\beta$ and $a, a'$ are $O(1)$ coefficients which are related to the partition function of the canonical ensemble. 
We show an explicit calculation in the following sections. 
The results for $n\geqslant 4$ are similar-looking expressions.
How these expression should be understood physically was already explained in \cite{Fujita:2017pju} and we summarise it in Sec. \ref{66}

%{\it with similar looking expressions for $n\geqslant 4$.}

The implication of the above statement is clear: { the $n$-th Renyi entanglement entropy can universally be decided up to several parameters}, which can be fitted with experimental/numerical data afterwards. We present their derivation in Sec. \ref{Calc Renyi}. {\it This completes the proof of the formula we presented in \cite{Fujita:2017pju}.}
The meaning of terms in each expression is also obvious --
the first terms are an \textit{offset}, the second ones mean a \textit{volume-law} (the slope being an effective dimension), and the third ones are a \textit{deviation} from it. 
Regarding the second Renyi entropy, \eqref{2ndRenyiMain} is simple. 
The second term indicates that the entanglement entropy grows linearly in terms of $\ell$ up to $\ell \sim L/2$. 
At $\ell=L/2$, especially, this deviation becomes $\ln 2$ for the second Renyi entropy, independent of the inverse temperature $\beta$ or the Hamiltonian.
We would like to stress that this fact is only peculiar to the second Renyi entropy, and generically the deviation at the center does depend on $\beta$ for the Renyi index greater than $2$. This can be a favourable fact in actually fitting the second Renyi entropy with the fit function above. 

We also obtain a result of von Neumann entropy of the cTPQ state for $e^\ell \ll e^L$. $\beta$-expansion of von Neumann entropy is written as
\begin{align}
S_{n \rightarrow 1} \simeq
	S_{\rm thermal}
	-\left(
	{1 \over 2} \frac{Z_B(2\beta)}{Z_B(\beta)^2}
	\sum_{r=1}^{\infty} {\beta^r \over Z_A(\beta)} \sum_{q =0}^{r} 
		{Z_A^{(q)}(0) Z_A^{(r-q)}(0) \over q ! (r-q)!} 
		\sum_{a=0}^q (-1)^{q-a} 
		\binom{q}{a} 
		B_{r-a}
	\right) +\ln R_*, 
	\label{Svn main}
\end{align}
%Here, we assume that we can ignore the contribution from the boundary, i.e., 
where 
\begin{align}
S_{\rm thermal} 
&\equiv 
 \beta \left(
	\langle H_A \rangle - F_A(\beta) 
	\right), 
\end{align}
$F_A(\beta) \equiv 
{1\over \beta}\ln\left({Z_A(\beta)} \right) $ 
is the free energy, 
$\langle H_A \rangle \equiv \beta {Z_A^{(1)}(\beta) \over Z_A(\beta)} $ 
is an average energy at the inverse temperature $\beta$, 
and 
\begin{align}
	 R_* \equiv \lim_{n \rightarrow 1} {\ln R_n(\beta) \over n-1}.
\end{align}
We present its derivation in Sec.\ref{Calc vN}.
Although \eqref{Svn main} is complicated, its 2nd term is $e^{-O(L-2\ell)}$. 
Namely, the entanglement entropy is almost equal to the thermodynamic entropy and the correction is exponentially small when $e^\ell \ll e^L$. 

\subsection{Calculation of the $n$-th Renyi entropy} \label{Calc Renyi}
%\subsubsection{Calculation of $\overline{\trA\rho_A^n }$}
\subsubsection{Diagrammatic representation of $\overline{\trA\rho_A^n }$}
\begin{figure}[htbp]
        \begin{center}
           \includegraphics[width=0.6\columnwidth]{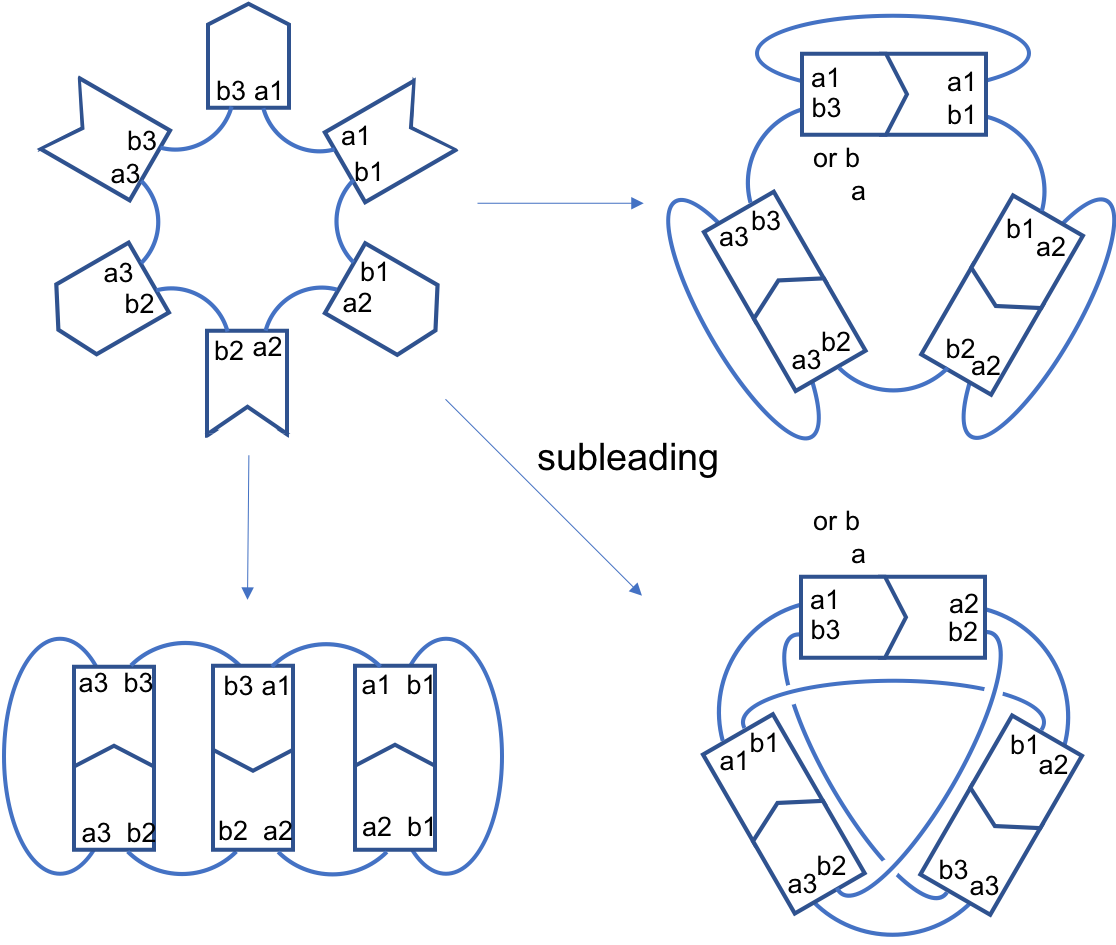}
    \caption{Fig. \ref{3}: All the $n=3$ graphs at leading order in large-$d$ and one sub-leading non-planer graph.}
    \label{3}
        \end{center}
        %\caption{Fig. \ref{2}: Ways to contract $n=3$ graph.}
    \end{figure}

By straightforward calculation, we get
\begin{align}
\trA\rho_A^n=\frac{1}{\left[\Tr\left(e^{-\beta H}\right)\right]^n}
\sum_{a_*^*,b_*^*}
\Bigl[
c_{a^1_1b^1_1}&c^*_{a^2_2b^2_1}
c_{a^1_2b^1_2}c^*_{a^2_3b^2_2}
\cdots c_{a^1_nb^1_n}c^*_{a^2_1b^2_n}\notag\\
&\times
\Braket{a^0_1,b^0_1|e^{-\beta H/2}|a^1_1,b^1_1}
\Braket{a_1^2,b_1^2|e^{-\beta H/2}|a^0_2,b^0_1}\notag\\
&\times\cdots\notag\\
&\times\Braket{a^0_n,b^0_n|e^{-\beta H/2}|a^1_n,b^1_n}
\Braket{a_n^2,b_n^2|e^{-\beta H/2}|a^0_1,b^0_n}
\Bigr].
\label{trcon}
\end{align}
We again represent this in terms of diagrams as follows:
\begin{equation}
\includegraphics[width=0.3\columnwidth]{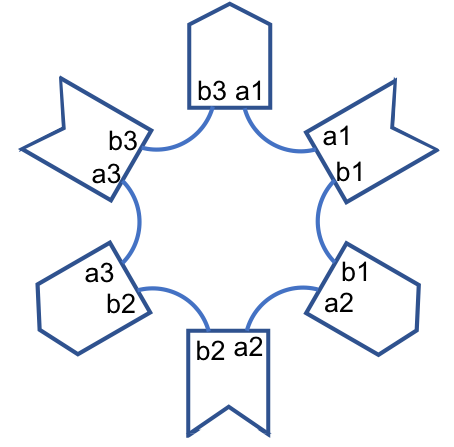}
\label{diagram finite}
\end{equation}
Here we represented $e^{-\beta H/2}\Ket{a_*^1,b_*^1}$ and
$\Bra{a_*^2,b_*^2}e^{-\beta H/2}$ as
$\includegraphics[height=1.5\baselineskip]{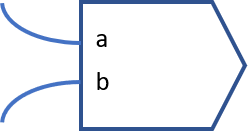}$ and $\includegraphics[height=1.5\baselineskip]{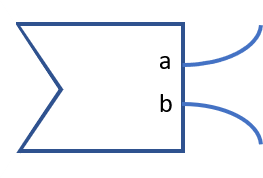}$, respectively.
These are connected with lines, which represent $\Bra{a_*^0,b_*^0}$ and $\Ket{a_*^0,b_*^0}$.
By taking an average over random variables, we
contract each box only once (Fig. \ref{3}) -- again as in Sec. \ref{nthrenyi}, contracting
twice will only count contributions which is subleading in $d$-scaling. 
Here, we represent the contraction as follows. 
\begin{align}
\vcenter{\hbox{\includegraphics[width=0.15\columnwidth]{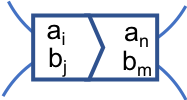}}}
=\bra{a_i^0, b_j^0} e^{-\beta H} \ket{a_n^0,b_m^0}	
\end{align}
In addition, the contribution which comes from
the diagrams which cannot be put on a plane,
\begin{equation}
\includegraphics[width=0.3\columnwidth]{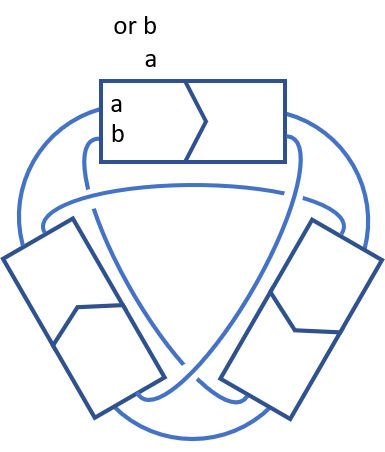}
\end{equation}
are subleading in $d$-scaling 
when $\beta$ is $O(1)$
because those graphs would lack the number of traces in the limit $\beta\to 0$.

\subsubsection{Relating new diagrams with the old ones}

The new graph \eqref{diagram finite} that we invented above have a correspondence
with the old one \eqref{npolygon} invented for the random spin system.
If we only consider diagrams which are leading in large-$d$ scaling,
the correspondence between 
the new and the old ones is one-to-one and is
as follows:
\begin{equation}
\includegraphics[width=0.3\columnwidth]{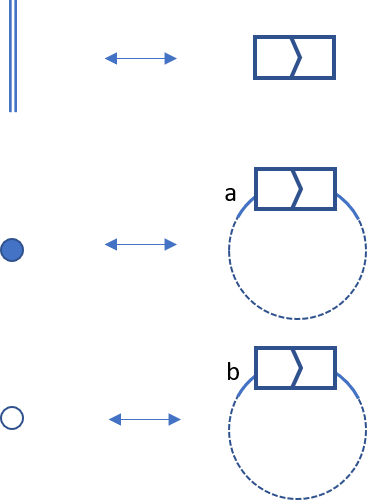}
\end{equation}
%Describing in texts, the boxes are replaced with double lines, while loops are replaced with blue or white dots depending on which subscript
%they have.
%Examples of this replacement are shown in the equations below.
%We are, hereafter, going to resort mostly to the old diagrams
%when we represent various trace contributions to (\ref{trcon})
%that are leading order in large-$d$.

\subsubsection{Calculation of $\overline{\trA\rho_A^n }$}

Calculation of  $\overline{\trA\rho_A^n}=\sum_{\text{All}}(\text{diagrams})$
is done in a same manner
as in Sec. \ref{calcrand},
but the actual calculation for generic $n$ is much harder, or virtually impossible.
Given a concrete value of $n$, however,
it is possible to calculate the Renyi entropy with that particular index.
We are going to calculate the second and the third Renyi entropies
as examples.

\paragraph{(a) Second Renyi entropy}
The second Renyi entropy is
\begin{align}
S_2&=
\vcenter{\hbox{\includegraphics[width=0.5\columnwidth]{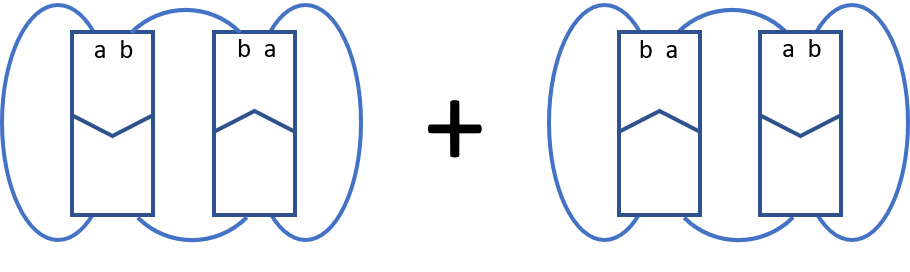}}}\\
%&=\vcenter{\hbox{\includegraphics[width=0.5\columnwidth]{6-1.png}}}\\
&=-\ln\left[
\frac{\trA{(\trB(e^{-\beta H})^2)}+\trB{(\trA(e^{-\beta H})^2)}}{(\Tr{e^{-\beta H})^2}}
\right]\label{TPQ2ndRenyi}
\end{align}

\paragraph{(b) Third Renyi entropy}
The third Renyi entropy is
\begin{align}
S_3&=\frac{1}{3-1}\times\left(
\vcenter{\hbox{\includegraphics[width=0.7\columnwidth]{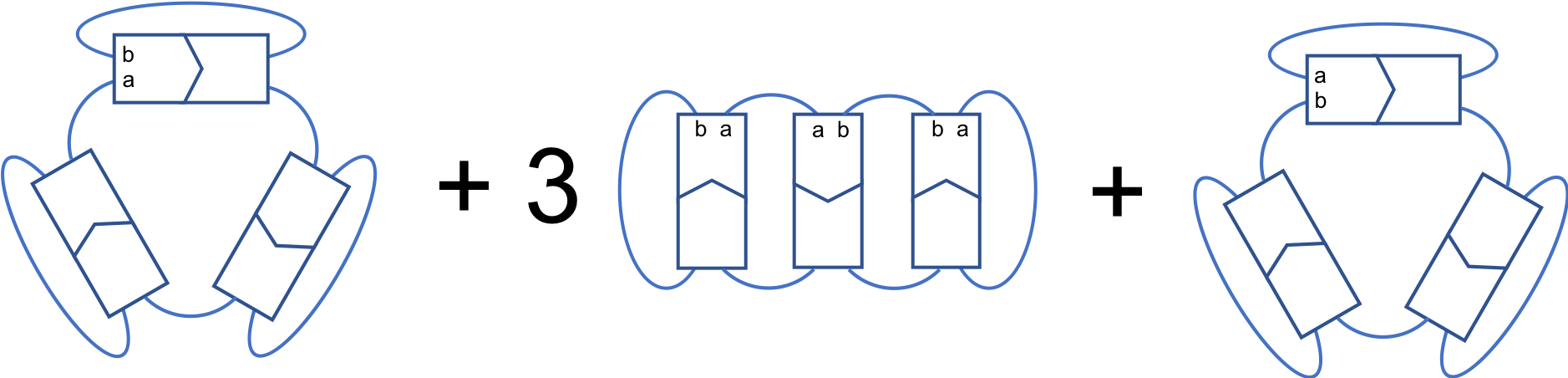}}}
\right)\\
%&=
%\frac{1}{2}\times\left(
%\vcenter{\hbox{\includegraphics[width=0.7\columnwidth]{8.png}}}
%\right)
%\\
&=-\frac{1}{2}\ln\left[
\frac{{
\trA{(\trB(e^{-\beta H})^3)}}{+
3\Tr M
+\trB{(\trA(e^{-\beta H})^3)}}
}{(\Tr{e^{-\beta H})^3}}
\right],
\label{TPQ3rdRenyi}
\end{align}
where
\begin{equation}
M=
e^{-\beta H}\left(\trB(e^{-\beta H})\otimes\trA(e^{-\beta H})\right)
\end{equation}

\subsection{Universality among Renyi entanglement entropies} \label{universality}

We are going to try to simplify the above result using the boundedness and the translation-invariance of the Hamiltonian and extensivity of the free energy.
This is done in two steps.
The implication of the resulting expression is essential -- { the Renyi entanglement entropy can be determined by finite unknown parameters of order $O(1)$}, as promised in the introduction.

\subsubsection{First step: rewriting each term with respect to the partition function}
As we assume that the interaction of the Hamiltonian is bounded, we can split the Hamiltonian into one in subsystem $A$, one in $B$, and one including interactions in $A$ and $B$:
\begin{equation}
H=H_{A}+H_{B}+H_{\rm int}.
\end{equation}
By using this decomposition, it is possible, at leading order in large-$d_{A,B}$, to replace each of the terms in the
$n$-th Renyi entropy using $Z_{A,B}(\beta)\equiv \mathop{\Tr_{A,B}} (e^{-\beta H_{A,B}})$ and several unknown $O(1)$ parameters, $P(\beta)$, $Q(\beta)$, etc., coming from the boundary term, $H_{\rm int}$.\footnote{This fact can be derived from the existence of transfer matrices. Also, be careful about the fact that those parameters are all dependent on $\beta$, although we will refer to them as ``parameters''.} 
We are listing some of the examples of this type of decomposition below
\begin{align}
\trA{\left(\trB(e^{-\beta H})^2\right)}&=P(\beta)\times Z_{A}(2\beta)\times Z_{B}(\beta)^2\\
\trB{\left(\trA(e^{-\beta H})^2\right)}&=P(\beta)\times Z_{A}(\beta)^2\times Z_{A}(2\beta)\\
\Tr{\left(e^{-\beta H}\right)}&=Q(\beta)\times Z_{A}(\beta)\times Z_{B}(\beta).
\end{align}

Now \eqref{TPQ2ndRenyi} becomes 
\begin{equation}\label{Z2ndRenyi}
S_2=-\ln R(\beta)-\ln\left(\frac{Z_A(2\beta)}{Z_A(\beta)^2}+\frac{Z_B(2\beta)}{Z_B(\beta)^2}\right),
\end{equation}
where $R(\beta)\equiv P/Q$.
Likewise, \eqref{TPQ3rdRenyi} becomes 
\begin{equation}\label{Z3rdRenyi}
S_3=-\ln R^{\prime}(\beta)-\ln\left(\frac{Z_A(3\beta)}{Z_A(\beta)^3}+
3\times\frac{Z_A(2\beta)}{Z_A(\beta)^2}\frac{Z_B(2\beta)}{Z_B(\beta)^2}
+\frac{Z_B(3\beta)}{Z_B(\beta)^3}\right),
\end{equation}
where $R^{\prime}(\beta)$ is again an unknown $O(1)$ parameter coming from the boundary terms.
The above procedure is just in the spirit of Suzuki-Trotter decomposition \cite{10.2307/2033649, Suzuki1976}.

\subsubsection{Second step: using extensivity}

Extensivity of the free energy lets us even simplify the expression for the Renyi entropies.
We here list two examples of extensivity relations that is of use in simplifying $S_2$. Note that again these are only true at leading order in large-$d_A$ and $d_B$:
\begin{align}
\frac{Z_A(2\beta)}{Z_A(\beta)^2}&=S(\beta)\times a(\beta)^{-\ell} \\
\frac{Z_B(2\beta)}{Z_B(\beta)^2}&=S(\beta)\times a(\beta)^{-L+\ell},
\end{align}
where $S(\beta)$ and $a(\beta)$ are, as usual, unknown $O(1)$ parameters coming from the details of the theory. 
Note that the inequality $1< a(\beta)\leqslant 2$ 
holds because of the concavity and monotonicity of the free energy (the equality holds when $\beta=0$).
By using those relations, \eqref{TPQ2ndRenyi} becomes
\begin{equation}
S_2=-\ln K(\beta)+\ell\ln a(\beta)-\ln\left(1+a(\beta)^{-L+2\ell}\right),
\end{equation}
where $K(\beta)\equiv S(\beta)R(\beta)$.\footnote{
Note that this expression is symmetric under inversion at $\ell=L/2$ as it should be.
}
This recovers the result for the random spin system at $\beta=0$.
Likewise, \eqref{TPQ3rdRenyi} becomes
\begin{equation}
S_3=-\ln K_2^{\prime}(\beta)+{1\over 2}\ell\ln b(\beta)-{1\over 2}\ln\left(1+K_1'(\beta){b(\beta)^\ell \over a(\beta)^{L}}+b^{\prime}(\beta)^{-L+2\ell}\right)
%S_3=-\ln K^{\prime}(\beta)+\ell\ln a^{\prime}(\beta)-\ln\left(1+T(\beta)a(\beta)^{-L}a^{\prime}(\beta)^\ell+a^{\prime}(\beta)^{-L+2\ell}\right),
\end{equation}
again consistent with the already derived expression for the random spin system at $\beta=0$.

\subsubsection{More universality in the thermodynamic limit}

%{\it This result was not presented in \cite{Fujita:2017pju}.}

As we mentioned in the last subsection, the deviation from a volume-law at the middle is generically dependent on the temperature. 
This, denoted $\varDelta S_n(L/2)$, is schematically written as
\begin{equation}
\varDelta S_n(L/2)=\ln\left(
1+\sum_i T_i(\beta)\times c_i(\beta)^{-L/2}+(\#)^{0}
\right)
=\ln\left(
2+\sum_i T_i(\beta)\times c_i(\beta)^{-L/2}
\right)
\end{equation}
Again because of the concavity and the monotonicity of the free energy, we have 
$c_i(\beta)>1$.
Hence, as you approach the thermodynamic limit, or when you take $L$ large,
the deviation of the general $n$-th Renyi entropy from a volume-law at $\ell=L/2$ quickly approaches
$\ln 2$ for any $\beta>0$, again independent of the inverse temperature $\beta$ or the details of the model.

\subsection{von Neumann entanglement entropy in finite temperature systems}\label{Calc vN}

%{\it This result was not presented in \cite{Fujita:2017pju}.}

Although it seems as if a very hard task to derive the expression for the
$n$-th Renyi entropy and hence the von Neumann entropy at finite $\beta$ as a result of taking a limit of $n\to 1$, it is nevertheless possible to derive the general expression if you wish to expand in terms of $\beta$.
We consider the case where $Z_B(\beta ) \gg Z_A(\beta ) $ 
(The readers are also referred to \cite{HUANG2018}, where the result is for any subsystem sizes, $A$ and $B$).

Let us return to Eq.~(\ref{Z2ndRenyi}) and (\ref{Z3rdRenyi}). $S_2$ is
\begin{equation}
S_2=\ln \left( \frac{Z_A(\beta)^2}{Z_A(2\beta)} \right)
-\ln\left(1+\frac{Z_A(\beta)^2}{Z_A(2\beta)}\frac{Z_B(2\beta)}{Z_B(\beta)^2}\right) -\ln R(\beta), 
\end{equation}
and $S_3$ is simplified as 
\begin{equation}
S_3 \simeq
{1\over 2}\ln\left(\frac{Z_A(\beta)^3}{Z_A(3\beta)}\right) 
-{1\over 2}\ln\left(1+
3\frac{Z_A(\beta)Z_A(2\beta)}{Z_A(3\beta)}\frac{Z_B(2\beta)}{Z_B(\beta)^2}
\right) 
-\ln R^{\prime}(\beta),
\end{equation}
where $R^{\prime}(\beta)$ and $R^{\prime}(\beta)$ are some constants.
In the similar manners, we can obtain the simple expression of the Renyi entropies when $Z_B(\beta ) \gg Z_A(\beta )$:
\begin{align}
S_n \simeq
{1\over n-1}\ln\left(\frac{Z_A(\beta)^n}{Z_A(n\beta)}\right) 
-{1\over n-1}\ln\left(1+
{n \over 2} \frac{Z_B(2\beta)}{Z_B(\beta)^2}
\sum_{k,m \leq 1, k+m=n}\frac{Z_A(k\beta) Z_A(k\beta)}{Z_A(n\beta)}
\right) 
-\ln R_n(\beta)
\end{align}
where $R_n(\beta)$ is a constant of $O(1)$. 
In order to take the analytic continuation, we expand $Z_A(k\beta)$ and $Z_A(m\beta)$ in terms of $\beta$. Then, the $O(\beta^r)$ terms are 
\begin{align}
	\sum_{q=0}^{r}
	Z_A^{(q)}(0) Z_A^{(r-q)}(0)
	{1\over q ! (r-q)!} 
	\sum_{k=1}^{n-1}
	k^{r-q} (n-k)^{q},
\end{align}
where 
\begin{align}
	Z_A^{(q)}(0) 
	\equiv
	\left. {\partial Z_A(x) \over \partial x}\right|_{x=0}.
\end{align}
We thus further expand the summation
\begin{align}
	\sum_{k=1}^{n-1} k^{r-q} (n-k)^{q}
	&= \sum_{k=1}^{n-1}\sum_{a=0}^q (-1)^{q-a} n^a 
		\binom{q}{a} 		k^{r-a} \\
	&= \sum_{a=0}^q (-1)^{q-a} n^a 
		\binom{q}{a} 		
		\frac{(n-1+B)^{r-a+1}-B^{r-a+1}}{r-a+1} \label{Sn Bernoulli}
\end{align}
where $B$ is the Bernoulli number in the umbral form. 
Namely, suppose $B_j$ is the Bernoulli number, one formally treats the indices $j$ in a sequence $B_j$
as if they were exponents. 
For example, in the umbral form we can write
\begin{align}
\sum_{k=1}^n \binom{n}{k} B_n = (1 + B)^n
\end{align}
In Eq.~(\ref{Sn Bernoulli}), it is possible to take $n \rightarrow 1$ limit and we get 
\begin{align}
	\lim_{n \rightarrow 1} {1 \over n-1}
	\sum_{k=1}^{n-1} k^{r-q} (n-k)^{q}
	&= \sum_{a=0}^q (-1)^{q-a} 
		\binom{q}{a} 
		B_{r-a}
\end{align}
Therefore, we obtain the $\beta$-expansion of von Neumann entropy of the TPQ state:
\begin{align}
S_{n \rightarrow 1} \simeq
	S_{\rm thermal}
	-\left(
	{1 \over 2} \frac{Z_B(2\beta)}{Z_B(\beta)^2}
	\sum_{r=1}^{\infty} {\beta^r \over Z_A(\beta)} \sum_{q =0}^{r} 
		{Z_A^{(q)}(0) Z_A^{(r-q)}(0) \over q ! (r-q)!} 
		\sum_{a=0}^q (-1)^{q-a} 
		\binom{q}{a} 
		B_{r-a}
	\right) +\ln R_*, \label{S1_beta}
\end{align}
%Here, we assume that we can ignore the contribution from the boundary, i.e., 
where 
\begin{align}
S_{\rm thermal} 
&\equiv 
 \beta \left(
	\langle H_A \rangle - F_A(\beta) 
	\right), 
\end{align}
$F_A(\beta) \equiv 
{1\over \beta}\ln\left({Z_A(\beta)} \right) $ 
is the free energy, 
$\langle H_A \rangle \equiv \beta {Z_A^{(1)}(\beta) \over Z_A(\beta)} $ 
is the average energy at the inverse temperature $\beta$, 
and 
\begin{align}
	 R_* \equiv \lim_{n \rightarrow 1} {\ln R_n(\beta) \over n-1}.
\end{align}

When $\beta = 0$, Eq.~(\ref{S1_beta}) reduces to
\begin{align}
S_{n \rightarrow 1} \simeq
	S_{\rm thermal}
	-\left(
	{1 \over 2} \frac{Z_A(0)}{Z_B(0)}
	\right),
\end{align}
which reproduces the result given by Page in this limit.

\section{Example: Ising model}
\label{Example}
As an illustration, we apply our formulation to the Ising model, 
and calculate the second Renyi entropy.
The results in this section support the validity of the approximations and our main results in the last section. 
We consider one-dimensional Ising model 
\begin{align}
H=\sum_{i=1}^{L-1} J \sigma_i^z \sigma_{i+1}^z + 
			\sum_{i=1}^{L}h \sigma_i^z, 
\end{align}
with the {\em open} boundary condition for the simplicity. 
Since the Hamiltonian is diagonal, the reduced density matrix of the TPQ state can be simplified. 
\begin{align}
	\rho_A
	={1 \over Z} \sum_{a_1,a_2,b_1} c_{a_1 b_1} c^*_{a_2 b_1} 
		e^{-{1\over 2}\beta \left\{ E(a_1 b_1) + E(a_2 b_1) \right\} }|a_1\rangle \langle a_2|
\end{align}
where $E(a_1 p_1) \equiv \langle a_1 p_1|H | a_1 p_1 \rangle$.
Therefore, we get
\begin{align}
	\Tr  \left[ \rho_A^n \right]
%	&=&{1 \over Z^n} \sum_{a_1,\cdots a_n,p_1, \cdots p_n} 
%	c_{a_1 p_1} c^*_{a_2 p_1} 
%	c_{a_2 p_2} c^*_{a_3 p_2} 
%	\cdots
%	c_{a_n p_n} c^*_{a_1 p_n} 
%	e^{-{1\over 2}\beta \left\{ 
%		E(a_1 p_1) + E(a_2 p_1) 
%		+E(a_2 p_2) + E(a_3 p_2) 
%		+\cdots
%		+E(a_n p_n) + E(a_1 p_n) 
%	\right\} } \\
	={1 \over Z^n} \sum_{a_1,\cdots a_n,p_1, \cdots p_n}
	c_{1 1} c^*_{2 1} 
	c_{2 2} c^*_{3 2} 
	\cdots
	c_{n n} c^*_{1 n} 
	e^{-{1\over 2}\beta \left\{ 
		E(1 1) + E(2 1) 
		+E(2 2)  
		+\cdots
		+E(n n) + E(1 n) 
	\right\} }.
\end{align}
where we use abbreviations 
$c_{a_i b_j} = c_{ij}$ and $E(a_i b_j) = E(ij)$, 
and its average is 
\begin{align}
	\overline{\Tr  \left[ \rho_A^n \right]}
	={1 \over Z^n} \sum_{a_1,\cdots a_n,p_1, \cdots p_n}
	\overline{c_{1 1} c^*_{2 1} 
	c_{2 2} c^*_{3 2} 
	\cdots
	c_{n n} c^*_{1 n}} 
	e^{-{1\over 2}\beta \left\{ 
		E(1 1) + E(2 1) 
		+E(2 2)
		+\cdots
		+E(n n) + E(1 n) 
	\right\} }.
\label{Ising Tr^n}
\end{align}
When we take the random average, there are many ways to contract $c_{a p}$ and $c^*_{a p}$, and 
$\overline{\Tr  \left[ \rho_A^n \right]}$ is a sum of all the contractions. 

%\subsubsection{Second Renyi entropy}
At $n=2$, the r.h.s of Eq.~(\ref{Ising Tr^n}) consists of two terms: 
\begin{align}
	\overline{\Tr  \left[ \rho_A^2 \right]}
	={1 \over Z^2} 
	\left[\sum_{a_1,p_1, p_2}
	e^{-\beta \left\{ 
		E(1 1) 
		+E(1 2)  
	\right\} }
	+
	\sum_{a_1,a_2,p_1}
	e^{-\beta \left\{ 
		E(1 1)  
		+E(2 1) 
	\right\} }
	\right]. \ \ \ 
\label{Ising Tr^2}
\end{align}
Since the r.h.s of Eq.(\ref{Ising Tr^2}) is symmetric with respect to the subsystems A and B, we only consider the first term. 
\begin{align}
&\sum_{a_1, p_1, p_2}
	e^{-\beta \left\{ 
		E(1 1) 
		+E(1 2)  
	\right\} } \nonumber \\
%&=&
%	\sum_{a1,p1,p2}
%	\langle \sigma_1| T | \sigma_2 \rangle
%	\cdots
%	\langle \sigma_{\ell-1}| T | \sigma_{\ell} \rangle
%	\langle \sigma_{\ell}| T | \tau_{\ell+1} \rangle
%	\cdots
%	\langle \tau_{L-1}| T | \tau_L \rangle
%	\langle \sigma_{\ell}| T | \tau'_{\ell+1} \rangle
%	\cdots
%	\langle \tau'_{L-1}| T | \tau'_L \rangle\\
=&
	\sum_{a1,p1,p2}
	\langle \sigma_1| T_2^{\ell-1} | \sigma_{\ell} \rangle
	\langle \sigma_{\ell}| T_1 | \tau_{\ell+1} \rangle
	\langle \tau_{\ell+1}| T_1^{L-\ell-2} | \tau_L \rangle
	\langle \sigma_{\ell}| T_1 | \tau'_{\ell+1} \rangle
	\langle \tau'_{\ell+1}| T_1^{L-\ell-2} | \tau'_L \rangle
\label{Ising 11, 12}
\end{align}
where $\{\sigma_i\}_i$, $\{\tau_i\}_i$, and $\{\tau'_i\}_i$ are 
a set of spins of $a_1$, $p_1$, and $p_2$, respectively, 
and $T_m$ is a transfer matrix s.t. 
\begin{align}
\langle \sigma_i| T_m | \sigma_{i+1} \rangle 
={\rm exp} \left[ m\beta \left(J\sigma_i \sigma_{i+1} 
		+ h {\sigma_i +\sigma_{i+1} \over 2} \right) \right] .
\end{align}

A further analysis is made by using the eigenvalues and the eigenvectors of $T_m$. 
$\lambda_\pm$ and $a_{\pm} |\uparrow \rangle+ b_{\pm} |\downarrow \rangle$ 
are the two eigenvalues ($|\lambda_+|>|\lambda_-|$) and eigenvectors of $T_1$, respectively, 
and
$\chi_\pm$ and $c_{\pm} |\uparrow \rangle+ d_{\pm} |\downarrow \rangle$ 
are the two eigenvalues and eigenvectors of $T_2$, respectively. 
Then, the partition function of this system is 
\begin{align}
Z(l, \beta) 
	\equiv&
	\sum_{\sigma_1, \sigma_\ell = \pm 1}\langle \sigma |T^{\ell-1}|\sigma_\ell \rangle \\
=& 
	(a_+^2 \lambda_+^{\ell-1}+a_-^2 \lambda_-^{\ell-1})
	+(a_+ b_+ \lambda_+^{\ell-1}+a_- b_-\lambda_-^{\ell-1})\\
	&+(b_+ a_+ \lambda_+^{\ell-1}+b_- a_-\lambda_-^{\ell-1})
	+(b_+^2 \lambda_+^{\ell-1}+b_-^2 \lambda_-^{\ell-1}) \\
=&
	(a_++b_+)^2\lambda_+^{\ell-1}+(a_-+b_-)^2 \lambda_-^{\ell-1})
\end{align}

Using these results, we calculate Eq.(\ref{Ising 11, 12}) 
\begin{align}
&\sum_{a_1, p_1, p_2}
	e^{-\beta \left\{ 
		E(1 1) 
		+E(1 2)  
	\right\} } \nonumber \\
&=
	(c_+(c_+ + d_+)\chi_+^{\ell-1} + c_-(c_- + d_-)\chi_-^{\ell-1})
	(a_+(a_+ + b_+)\lambda_+^{L-\ell} + a_-(a_- + b_-)\lambda_-^{L- \ell})^2 \nonumber \\
&+ (d_+(c_+ + d_+)\chi_+^{\ell-1} + d_-(c_- + d_-)\chi_-^{\ell-1})
	(b_+(a_+ + b_+)\lambda_+^{L-\ell} + b_-(a_- + b_-)\lambda_-^{L- \ell})^2 \label{Ising abcd}
\end{align}
When $L \geq \ell >> 1$, we drop the terms of $\lambda_-$ and $\chi_-$ in Eq.~(\ref{Ising abcd}) and get 
\begin{align}
&\sum_{a_1, p_1, p_2}
	e^{-\beta \left\{ 
		E(1 1) 
		+E(1 2)  
	\right\} } \nonumber \\
&=
	(c_+(c_+ + d_+)\chi_+^{\ell-1} )
	(a_+(a_+ + b_+)\lambda_+^{L-\ell} )^2
	+ (d_+(c_+ + d_+)\chi_+^{\ell-1})
	(b_+(a_+ + b_+)\lambda_+^{L-\ell})^2 \\
&=
	(a_+ + b_+)^2 (c_+ + d_+)(a_+^2 c_+ + b_+^2 d_+) 
	\chi_+^{\ell-1} \lambda_+^{2(L-\ell)} 
\end{align}
Hence, Eq.~(\ref{Ising Tr^2}) is 
\begin{align}
	\overline{{\rm Tr}[\rho_A^2]}
	&\simeq
	\frac{
	(a_+ + b_+)^2 (c_+ + d_+)(a_+^2 c_+ + b_+^2 d_+) 
	(\chi_+^{\ell-1} \lambda_+^{2(L-\ell)} 
	+\chi_+^{L-\ell-1} \lambda_+^{2\ell} )
	}
	{(a_+ + b_+)^4 \lambda_+^{2(L-1)}} \nonumber \\
	&=
	\left( \chi_+ \over \lambda_+^2 \right)^{\ell-1}
	\left( 
	1 + 	\left( \chi_+ \over \lambda_+^2 \right)^{L-2\ell}
	\right)
	\frac{
	(c_+ + d_+)(a_+^2 c_+ + b_+^2 d_+) 
	}
	{(a_+ + b_+)^2 } 
	,
\label{L,ell>>1}
\end{align}
and the second Renyi entropy is 
\begin{align}
	\overline{S_2}
	\simeq
	\ell \ln \alpha 
	-\ln \left( 
	1 + {1 \over \alpha ^{L-2\ell} }
	\right)
	+
	\left(\ln\frac
	{(a_+ + b_+)^2 }
	{
	(c_+ + d_+)(a_+^2 c_+ + b_+^2 d_+) 
	}
	-\ln \alpha
	\right). 
	\label{Ising Renyi2}
\end{align} 
where $\alpha \equiv {\lambda_+^2 \over \chi_+}$.
\if0
Moreover, it decays exponentially fast as $\ell$ gets small; 
\begin{align}
	-\ln \left( 
	1 + 	{1\over \alpha^{L-2\ell}}
	\right)
	\simeq
	- 	{1\over \alpha^{L-2\ell}}
\end{align}
where ${1\over \alpha^{L-2\ell}}$ is small.
\fi
%To see it more concretely, we will show some results for $h=0$.
In particular, when $h=0$
\begin{align}
	&\lambda_\pm =e^{\beta J} \pm e^{-\beta J} \\
	&a_{\pm} = c_{\pm} = {1\over \sqrt{2}} \\
	&b_{\pm} = d_{\pm} = \pm{1\over \sqrt{2}}. 
\end{align}
Thus, the third term in Eq.(\ref{Ising Renyi2}) is simplified. 
\begin{align}
	\ln\frac
	{(a_+ + b_+)^2 }
	{
	(c_+ + d_+)(a_+^2 c_+ + b_+^2 d_+) 
	}
	-\ln \alpha
	=
	\ln {2 \over \alpha}. 
\end{align}

The final result (\ref{Ising Renyi2}) consist of three terms, the volume-law slope, the deviation from it, and the offset term.
The 1st term gives a volume-law contribution. 
The 2nd term gives the deviation from the volume-law, and it takes a minimum value $-\ln 2$ at $\ell = {L \over 2}$. 
The 3rd term is the offset term because it is independent of $\ell$. 
The most important observation of this example is that Eq.~(\ref{Ising Renyi2}) perfectly recovers Eq.(\ref{2ndRenyiMain}). 
%This feature is consistent with Sec.~\ref{Main Result}. 
In contrast to Eq.~(\ref{2ndRenyiMain}), which is obtained by imposing a few assumptions, we do not assume anything to derive Eq.~(\ref{Ising Renyi2}) in this section. 
Hence, the results in this section support the validity of the assumptions in Sec.~\ref{universality}. 

\if0
In another case where $L >> 1$ but $\ell$ is small, $\lambda_-^{L-\ell}$ and $\chi_-^{L-\ell}$ in Eq.~(\ref{Ising abcd}) are negligible, and we obtain 
\begin{align}
&\sum_{a_1, p_1, p_2}
	e^{-\beta \left\{ 
		E(1 1) 
		+E(1 2)  
	\right\} } \nonumber \\
	&	(c_+(c_+ + d_+)\chi_+^{\ell-1} 
		+ c_-(c_- + d_-)\chi_-^{\ell-1})
	(a_+(a_+ + b_+)\lambda_+^{L-\ell} )^2 \nonumber \\
	&+ (d_+(c_+ + d_+)\chi_+^{\ell-1} 
		+ d_-(c_- + d_-)\chi_-^{\ell-1})
	(b_+(a_+ + b_+)\lambda_+^{L-\ell})^2 \\
&=
	(a_+ + b_+)^2 
	\left( (c_+ + d_+)(a_+^2 c_+ + b_+^2 d_+) \chi_+^{\ell-1} 
		+(c_- + d_-)(a_+^2 c_- + b_+^2 d_-) \chi_-^{\ell-1} 
	\right)
	\lambda_+^{2(L-\ell)} 
\end{align}
This modification from Eq.~(\ref{L,ell>>1}) brings some correction to Eq.~(\ref{Ising Renyi2}), although I didn't go into the detail of the correction here.

\section{Renyi-n Entropy}
\fi

\section{Physical understanding and Applications}
\label{66}
For numerical evidences and physical understanding of our formulae, the readers are referred to our previous work \cite{Fujita:2017pju}.
To summarise the paper, the formula works quite well for non-integrable models while not for integrable models.
This was attributed to the fact that our derivation only works well for fast-scrambling systems,
and hence we concluded that the formula in turn works as a diagnosis for chaotic systems.
We also checked that the formula fits well for states after a quantum quench after time-averaging, and the fit worked well for integrable as well as
non-integrable models. 

The applications of our formula could be wide-ranging.
Aside from the above mentioned diagnosis for chaotic systems, it was also proposed in \cite{Fujita:2017pju}
that it could detect ETH-MBL transitions with better accuracy.

Note that these could not have been achieved using conventional thermodynamics using Gibbs ensembles -- the states we consider (which can be experimentally realised too) are all pure states, and they would not have at all reproduced what we have computed so far.
Especially the $O(1)$ deviation in the middle is where the effect of pure states comes in directly, which again could be measured by experiments.

\section{Conclusion and Outlook}

We have derived the formula for the von Neumann/Renyi Page curves in a finite volume system.
We first computed the Renyi Page curves for the infinite temperature systems using a diagrammatic approach, and then
analytically continued to get the von Neumann Page curve, reproducing the result of Page \cite{Page:1993df}.

We then expanded the result to general interacting finite-temperature systems by using cTPQ states, 
and computed the Renyi Page curve using a similar diagrammatic technique.
We then explicitly showed the universality of the form of the Page curves using a finite number of thermodynamic constants,
from which we infer the effectiveness of the formulas in fitting with numerical or experimental data.
We also computed the von Neumann Page curve by using the high-temperature expansion.

There are a number of interesting directions to pursue in the future.
As was promoted in our previous work, \cite{Fujita:2017pju}, this formula is conjectured to be a diagnosis for fast-scrambled systems,
which might compliment the tedious task of computing the OTOC.
It would be interesting to collect evidences in this direction by numerics or experiments.
The advantage of this formula is that it works well for fast-scrambled models even at system sizes $L\sim 15$, and such computations for verifying our formula might be easier to come by than other formulas about entanglement.

It would be also intriguing to derive the von Neumann Page curve for $\beta=O(1)$.
Because the volume-law of Renyi entropies are not actually exact in large total volume limit, and becomes concave rather than convex \cite{PhysRevX.8.021026},
our formula surely only applies to the regime where $L\lesssim 30$.
Although by computational or experimental difficulty, this is by no means a practical problem,
it would be better to derive a complete formula for the von Neumann Page curve, which is known not to have this issue.

\section*{{Acknowledgement}}
The authors thank 
%J.~Eisert, F.~Pollmann, T.~Grover, 
M.~Oshikawa, 
T.~Sagawa, T.~Takayanagi, and H.~Tasaki for their valuable discussions. 
%The authors thank T.~N.~Ikeda and K.~Kawaguchi for their in-depth readings of the manuscript. 
The authors gratefully acknowledge the hospitality of the Yukawa Institute for Theoretical Physics at Kyoto University,
where this work was initiated during the long-term workshop YITP-T-16-01
``Quantum Information in String Theory and Many-body Systems''.
HF, YON and SS also gratefully acknowledge the hospitality of the Kavli Institute for Theoretical Physics,
where this work was improved and supported in part by the National Science Foundation under Grant No. NSF PHY-1125915.
The authors are supported by JSPS KAKENHI Grant Numbers
JP16J04752, JP16J01135, JP15J11250 and JP16J01143, 
and by World Premier International Research Center Initiative (WPI), MEXT, Japan.
HF and YON also acknowledge support from the ALPS program
and MW from the FMSP program, both under ``The Program for Leading Graduate Schools'' of JSPS.

\appendix

\section{Averaging random variables}
\label{ssec:rmt}
We rely on the work \cite{ULLAH196465} for averaging
random variables in Gaussian unitary ensemble (GUE).
Although the work above mostly calculate
the average of various random variables in
Gaussian orthogonal ensemble (GOE),
the generalisation to GUE is straightforward and
we will just show the result of the averaging below:
\begin{equation}
\underbrace{\overline{|c_{*,*}|^2\cdots|c_{*,*}|^2}}_{\text{$n$ times}}=\frac{1}{d(d-1)\cdots (d-n+1)}\sim \frac{1}{d^n}
\end{equation}
Other combinations just vanish at leading order in $1/d$.

\section{Deriving the von Neumann Page curve for the random spin system}

\subsection{von Neumann Page curve for the random spin system}
Here we analytically continue the Renyi Page curve to the von Neumann Page curve for the random spin system.
This requires the knowledge of the
Narayana polynomial \cite{2008arXiv0805.1274M}.
Narayana polynomial $\mathcal{N}_n(q)$ is defined as
\begin{align}
\label{nar}
\mathcal{N}_n(q)&=\sum_{k=1}^{n} N(n,k)q^{k-1}\\
&=q^{n-1}\sum_{k=1}^{n} N(n,k)\left(\frac{1}{q}\right)^{k-1}=q^{n-1}\mathcal{N}_n
\left(q^{-1}\right)
\label{inv}
\end{align}
and known to be represented in terms of Legendre polynomials as
\begin{align}
\mathcal{N}_n(q)&=
\frac{(q-1)^{n+1}}{q}\int\limits_0^{\frac{q}{q-1}}dx\, P_n(2x-1)\\
&=\frac{(q-1)^{n+1}}{(4n+2)q}\left[
P_{n+1}\left(\frac{q+1}{q-1}\right)-P_{n-1}\left(\frac{q+1}{q-1}\right)
\right]. \label{anacon}
\end{align}

Now let us analytic continue the function $S_n(\ell)$.
We work in a region where $0\muchlessthan\ell\leqslant L/2$, so
let us denote $d_A/d_B=q$, where $0\leqslant q\leqslant 1$. Then we have
\begin{equation}
S_n=\ell\ln 2-\frac{1}{n-1}\ln{\mathcal{N}_n(q)}
\label{Narayana2}
\end{equation}
For (\ref{anacon}) to be an analytic continuation of $\mathcal{N}_n(q)$,
note that $q$ has to satisfy $q\geqslant 1$, because of the presence of 
the term like $(q-1)^{n+1}$. This means in a region of
interest, $0\leqslant q\leqslant 1$, the expression (\ref{inv}),
rather than (\ref{nar}),
must be used alternatively in order to perform an analytic continuation
to $n=1$:
\begin{align}
\Delta(q)&\equiv \ell\ln 2- \lim_{n\to 1}S_n=\lim_{n\to 1}\frac{1}{n-1}\ln \left[q^{n-1}\mathcal{N}_n
\left(q^{-1}\right)\right]\\
&=
\left.\frac{\partial}{\partial n}\right|_{n=1}
\ln \left[q^{n-1}\mathcal{N}_n
\left(q^{-1}\right)\right]\\
&=
\left.\frac{\partial}{\partial n}\right|_{n=1}
\ln
\left[
\frac{(1-q)^{n+1}}{(4n+2)q}\left(
P_{n+1}\left(\frac{1+q}{1-q}\right)-P_{n-1}\left(\frac{1+q}{1-q}\right)
\right)
\right]\\
&=
\ln(1-q)-\frac{2}{3}+\frac{\left.\frac{\partial }{\partial\nu}\right|_{\nu=2}
P_\nu\left(\frac{1+q}{1-q}\right)
-\left.\frac{\partial}{\partial\nu}\right|_{\nu=0}P_\nu\left(\frac{1+q}{1-q}\right)
}{P_{2}\left(\frac{1+q}{1-q}\right)-P_{0}\left(\frac{1+q}{1-q}\right)}
\label{cumber}
\end{align}
This expression includes derivatives of Legendre polynomials in terms of their degrees.
These are known to be
\begin{equation}
\left.\frac{\partial P_\nu(z)}{\partial\nu}\right|_{\nu=n}=P_n(z)\ln\left(
\frac{z+1}{2}
\right)+R_n(z),
\end{equation}
where $R_n(z)$ is a certain polynomial of order $n$ \cite{0305-4470-39-49-006}.
Specifically, according to the paper above, we have
$R_0(z)=0$ and $R_2(z)=\frac{7}{4}z^2-\frac{3}{2}z-\frac{1}{4}$.
Plugging these into (\ref{cumber}),
we get
\begin{equation}
\Delta(q)=\frac{q}{2},
\end{equation}
so that the von Neumann entropy of the random spin system becomes
\begin{equation}
S=\ell\ln 2-\frac{1}{2}\frac{d_A}{d_B},
\label{pages}
\end{equation}
as promised.

\subsection{Infinite Renyi index limit of the random spin system}

Let us also take $n\to \infty$ in \eqref{nthrenyimaximal} to get the first eigenvalue of the reduced density matrix.
By using \eqref{Narayana2}
and \eqref{anacon},
we have
\begin{equation}
\Delta_{n}(q)=\frac{1}{n-1}\ln
\left[
\frac{(1-q)^{n+1}}{(4n+2)q}\left(
P_{n+1}\left(\frac{1+q}{1-q}\right)-P_{n-1}\left(\frac{1+q}{1-q}\right)
\right)
\right].
\end{equation}
Now, for large $n$, the asymptotic form of the Legendre polynomials can be found in \cite{olver2010nist, Temme:2015:AMfI}:
\begin{equation}
P_n(z)=\frac{1+\sqrt{1-z^{-2}}}{\sqrt{2\pi n\sqrt{1-z^{-2}}}}\left(\frac{1+\sqrt{1-z^{-2}}}{1-\sqrt{1-z^{-2}}}\right)^{n/2}+O(n^{-1}),
\end{equation}
where $z>1$.
By using this expression, the finite index limit of $\Delta_n(q)$ becomes
\begin{equation}
\lim_{n\to\infty}\Delta_{n}(q)
=\ln(1-q)+\ln\left[\frac{1+\sqrt{q}}{1-\sqrt{q}}\right] 
=2\ln\left[1+\sqrt{q}\right],
\end{equation}
and the min-entropy of the random spin system becomes
\begin{equation}
S_\infty=\ell\ln 2-2\ln\left[1+\sqrt{\frac{d_A}{d_B}}\right].
\end{equation}
Incidentally the maximal value of the min-entropy is
\begin{equation}
\frac{L}{2}\ln 2-2\ln 2,
\end{equation}
which can also be directly checked by taking $n\to\infty$ in \eqref{nthrenyimaximal}.

\section{log of average v.s. average of log}
\label{referee}
In this appendix we provide a proof of the following property:
\begin{equation}
\overline{\log \left[\trA \left(\rho_A^n\right)\right]}
=
\log\overline{\left[\trA \left(\rho_A^n\right)\right]}+O(1/d),
\label{whatwewant}
\end{equation}
where $d=\alpha^L$, $L$ is the system size, and $1<\alpha $
is the effective dimension of the system.
Note that $\alpha=2$ at infinite temperature for $S=1/2$ spin systems.
This fact is actually very intuitive, because at large-$d$, the variation
for $W[z,\bar{z}]\equiv \trA \left(\rho_A^n\right)$ is suppressed exponentially and
one should be able to replace the average of functions with functions of the average.

\subsection{The idea of the proof}

Let us set up the notations. We denote $W[z,\bar{z}]\equiv \trA \left(\rho_A^n\right)$, where $z$ is
the random complex number which we take averages over. 
We also write $\Omega\equiv\overline{W[z,\bar{z}]}$,
so we are going to prove
\begin{equation}
\overline{\log W[z,\bar{z}]}=\log\Omega+O(1/d)
\iff
\overline{\log\left[\frac{W[z,\bar{z}]}{\Omega}\right]}=O(1/d).
\end{equation}
Now we {\it formally} expand the $\log$ around $\frac{W[z,\bar{z}]}{\Omega}=1$ and we get the following,
\begin{equation}
\overline{\log\left[\frac{W[z,\bar{z}]}{\Omega}\right]}
= -\frac{1}{2}\overline{\left(\frac{W[z,\bar{z}]}{\Omega}-1\right)^2}+
\frac{1}{3}\overline{\left(\frac{W[z,\bar{z}]}{\Omega}-1\right)^3}
-\frac{1}{4}\overline{\left(\frac{W[z,\bar{z}]}{\Omega}-1\right)^4}
+\cdots.
\end{equation}
Note that we have used $\overline{\frac{W[z,\bar{z}]}{\Omega}-1}=0$.
%\if0
%Now then, it reduces to the discussion of the usual moments and cumulants.
%Notice that this expansion is just a formal expansion, because the expression can only converge for $0<W[z,\bar{z}]<2$.
%The justification of the whole discussion will be given in the next subsection, but let us use this expansion for a moment to get a feel of the idea of the proof.
%\fi

Let us discuss the first term $\overline{\left(\frac{W[z,\bar{z}]}{\Omega}-1\right)^2}$.
This gives
\begin{equation}
\overline{\left(\frac{W[z,\bar{z}]}{\Omega}-1\right)^2}
=\frac{\overline{W^2}-\Omega^2}{\Omega^2},
\end{equation}
but $\overline{W^2} -\Omega^2$ can be calculated to give $\Omega^2\times O(1/d)$.
%\if0
%and now $\overline{W^2}-\Omega^2$ can be calculated using the same diagrammatic method as before -- 
%$\overline{W^2}$ can be represented by contracting two $2n$-gon diagrams,
%but subtracting $\Omega^2$ removes all the disconnected pieces.
%Then we only have diagrams which gives the contribution that goes $\Omega^2\times O(1/d)$, so that we have
%\begin{equation}
%\overline{\left(\frac{W[z,\bar{z}]}{\Omega}-1\right)^2}=O(1/d)
%\end{equation}
%\fi
Likewise, we can see that the terms like $ \overline{\left(W-\Omega\right)^m}$ would only scale as $\Omega^m\times O(1/d^{\lfloor m/2\rfloor})$,
\begin{equation}
\frac{ \overline{\left(W-\Omega\right)^m}  }{\Omega^m}=O(1/d^{\lfloor m/2\rfloor}).
\label{trustme}
\end{equation}
By summing up all the contributions, we will get
\begin{equation}
\overline{\log\left[\frac{W[z,\bar{z}]}{\Omega}\right]}=\sum_{l=1}^\infty a_l d^{-l},
\end{equation}
where $a_l$ is independent of $d$ and scales exponentially as $l$ as seen from the direct computation.
Therefore, for sufficiently large $d \: (= O(e^L))$,
the right hand side of the above formula converges, which is of order $O(1/d)$.

\subsection{Proof}

The rigorous proof of \eqref{whatwewant} can be done using the idea above, but we still have to justify the expansion of the logarithm, because it can include the piece where the argument in the log is greater than 2, which is out of the convergence radius.
The rigorous proof, then, only includes the expression using the Taylor expansion up to a finite order and a remaining term.

Let us write the probability distribution of $\Phi = W[z,\bar{z}] / \Omega$ to be $P[\Phi]$,
so that we have
\begin{equation}
\overline{\log\left[\frac{W[z,\bar{z}]}{\Omega}\right]}
=\int_{1/d_A^{n-1} }^{d_A^{n-1}} d\Phi P[\Phi]\log \Phi.
\end{equation}
%We here took the integration range from $0$ to $\infty$, but by construction $1/d_A^{n-1}\leqslant W[z,\bar{z}]=\trA \left(\rho_A^n\right)\leqslant1$, so that the range can be reduced to $1/d_A^{n-1}\leqslant\Phi\leqslant d_A^{n-1}$ ($d_A = 2^{l_A}$ is the dimension of the subsystem $A$).
We here take the integration range from $1/d_A^{n-1}$ to $d_A^{n-1}$ since by construction
$1/d_A^{n-1}\leqslant W[z,\bar{z}]=\trA \left(\rho_A^n\right)\leqslant1$  and
$1/d_A^{n-1}\leqslant\Phi\leqslant d_A^{n-1}$, where $d_A $ is the dimension of the subsystem $A$.
Here we assume the subsystem $A$ is smaller than the rest of the system, $B=\overline{A}$ 
(when $A$ is larger than $B$ then the bound is given by $d_B$).
Now we expand $\log \Phi=(\Phi-1)-(\Phi-1)^2/(2\xi^2)$, where $\xi$ is in between $1$ and $\Phi$ (the Taylor theorem),
\begin{equation}
\overline{\log{\Phi}}=\int_{1/d_A^{n-1} }^{d_A^{n-1}} d\Phi P[\Phi](\Phi-1)
-\frac{1}{2} \int_{1/d_A^{n-1} }^{d_A^{n-1}} d\Phi P[\Phi] \frac{(\Phi-1)^2}{\xi^2},
\end{equation}
but the first term gives zero because $\overline{\Phi-1}=0$.
In the following we divide the range of integration into two parts,
$[1/d_A^{n-1},1/2]$ and $[1/2, d_A^{n-1}]$, and evaluate each of them, respectively.

\paragraph{Integration range $[1/d_A^{n-1},1/2]$}
We would like to evaluate
\begin{equation}
I_1\equiv\int_{1/d_A^{n-1}}^{1/2} d\Phi P[\Phi] \frac{(\Phi-1)^2}{\xi^2} \:\: \geq 0.
\end{equation}
Because $\xi>1/d_A^{n-1}$, we have
\begin{equation}
I_1
<
\int_{1/d_A^{n-1}}^{1/2} d\Phi P[\Phi] d_A^{2(n-1)}(\Phi-1)^2,
\end{equation}
and also because $(\Phi-1)^2<1$,
\begin{equation}
I_1
<
\int_{1/d_A^{n-1}}^{1/2} d\Phi P[\Phi] d_A^{2(n-1)}(\Phi-1)^2
<
\int_{1/d_A^{n-1}}^{1/2} d\Phi P[\Phi] d_A^{2(n-1)}.
\end{equation}
This quantity has an upper bound from the Chebyshev inequality for higher moments.
%(this was actually computed in the previous subsection).
The inequality on the $2n$-th moment tells that
$\mathrm{Prob} \left( | \Phi - 1 | > 1/2 \right) \leq 2^{2n} \overline{(\Phi-1)^{2n}}$
so we obtain
\begin{equation}
I_1 <
2^{2n}d_A^{2(n-1)}\times \overline{(\Phi-1)^{2n}}=O(1/d),
\end{equation}
where we have used Eq.~\eqref{trustme} and $d_A\leqslant d^{1/2}$.

\paragraph{Integration range $[1/2,d_A^{n-1}]$}
We would then like to evaluate
\begin{equation}
I_2\equiv\int^{d_A^{n-1}}_{1/2} d\Phi P[\Phi] \frac{(\Phi-1)^2}{\xi^2}  \:\: \geq 0.
\end{equation}
Because $\xi>1/2$ we have 
\begin{equation}
I_2
<
4\times \int^{d_A^{n-1}}_{1/2} d\Phi P[\Phi] {(\Phi-1)^2}.
\end{equation}
%Since the integrand $(\Phi-1)^2$ is non-negative, we can extend the range of integration:
Also,
\begin{equation}
I_2 <
4\times \int^{d_A^{n-1}}_{1/2} d\Phi P[\Phi] {(\Phi-1)^2}
<
4\times d_A^{2(n-1)} \int^{d_A^{n-1}}_{1/2} d\Phi P[\Phi]
=O(1/d),
\end{equation}
where the last inequality again comes from the result in the previous subsection.

\paragraph{Sum of the above two terms}
Summing up the above two results, we have
\begin{equation}
  \overline{\log\Phi}  = O(1/d),
\end{equation}
which is the desired result.

\subsubsection*{Comments on Eq.~\eqref{trustme} }
We have not given any proof of \eqref{trustme}, since proving this
in full generality is too complicated.
The proof goes the same as in deriving the result of the average of the R\'enyi entropy
(just contracting the indices in the random number $z$), and when $m=2$ and $n=2$ for example
we have 
\begin{equation}
\overline{\left({W[z,\bar{z}]}-\Omega\right)^2}
=\sum_{ijklmop} Z_{op}^{ij}Z_{kl}^{ij}Z_{kl}^{mn}Z_{op}^{mn}
+\trB\left[\trA Z^2 \left(\trA Z\right)^2\right]+(A\leftrightarrow B)
\end{equation}
where $Z\equiv e^{-\beta H}$, taking indices in the subspace $H_A$ (upper) and $H_B$ (lower), respectively.
By following the argument in the main text to pull out the extensive contributions,
one can see the terms in the right hand side divided by $\Omega ^2$ are of the order of  $O(1/d)$.
%Notice that in the context of the subsystem ETH, there can be higher order terms like $O(1/\alpha^{\ell^2})$, but this is still smaller than $O(1/d)$, so we omit them.

\bibliographystyle{JHEP} 	%which one do you like?
\bibliography{references}

\end{document}